\documentstyle[amstex,12pt]{article}

\renewcommand{\theequation}{\arabic{equation}}
\newcommand{\EQ}{\begin{equation}}

\newcommand{\EN}{\end{equation}}

\newcommand{\bear}{\begin{eqnarray}}
\newcommand{\ear}{\end{eqnarray}}

\begin{document}

\topmargin 0pt \oddsidemargin 5mm

\renewcommand{\thefootnote}{\fnsymbol{footnote}}

\newpage \setcounter{page}{0} 
\begin{titlepage}     
\begin{flushright}
UFSCar-HEP-06/15
\end{flushright}
\vspace{0.5cm}
\begin{center}
\large{Solutions of the reflection equation for the $U_q[G_2]$ vertex model.} \\
\vspace{1cm}
\vspace{1cm}
 {\large A. Lima-Santos and M.J. Martins } \\
\vspace{1cm}
\centerline{\em Departamento de F\'isica, Universidade Federal de S\~ao Carlos}
\centerline{\em Caixa Postal 676, 13565-905, S\~ao Carlos, Brazil}
\vspace{1.2cm}   
\end{center} 
\begin{abstract}
We investigate the possible regular solutions of the boundary Yang-Baxter equation for the
fundamental $U_q[G_2]$ vertex model. We find four distinct classes of 
reflection matrices such that half of them are diagonal while the other half are
non-diagonal. The latter are parameterized  by two continuous parameters 
but only one solution has all entries non-null.  
The non-diagonal solutions do not
reduce to diagonal ones at any special limit of the free-parameters.

\end{abstract}
\vspace{.2cm}
\centerline{Keywords: Reflection Equation, K-matrices, $U_q[G_2]$}
\vspace{.2cm}
\centerline{August 2006}
\end{titlepage}

\renewcommand{\thefootnote}{\arabic{footnote}}

\section{Introduction}

It is, by now, well known that integrable two-dimensional systems of
statistical mechanics can be derived from the Yang-Baxter equation \cite{BA}%
. The respective Boltzmann weights can be related in a natural manner to the
elements of a Yang-Baxter $\check{R}$-matrix solution $\check{R}_{ab}(x)\in
C_{a}^{N}\otimes C_{b}^{N}$ satisfying the relation, 
\begin{equation}
\check{R}_{12}(x)\check{R}_{23}(xy)\check{R}_{12}(y)=\check{R}_{23}(y)\check{%
R}_{12}(xy)\check{R}_{23}(x)  \label{YB}
\end{equation}%
for arbitrary spectral parameters $x$ and $y$.

For an open statistical system not all possible types of boundaries are
compatible with the Yang-Baxter condition (\ref{YB}). However, the
integrability at the boundary may be assured when the boundary weights $%
K(x)\in C^{N}$ fulfill the so-called reflection equation \cite{CK,SK}, which
reads 
\begin{equation}
\check{R}_{12}(x/y)\overset{1}{K}(x)\check{R}_{12}(xy)\overset{1}{K}(y)=%
\overset{1}{K}(y)\check{R}_{12}(xy)\overset{1}{K}(x)\check{R}_{12}(x/y)
\label{RE1}
\end{equation}%
where $\overset{1}{K}(x)=K(x)\otimes I_{N}$ and $I_{N}$ is the $N\times N$
identity matrix.

An important class of $\check{R}$-matrices are those based on a quantum
group $U_{q}[{\cal {G}]}$ $q$-deformation of a classical Lie algebra ${\cal {%
G}}$ \cite{JI}. For these models, from a given solution $K(x)$ of Eq.(\ref%
{RE1}) one can, in principle, construct families of commuting transfer-matrix 
\cite{SK,NEP}. This fact has motived several authors, see for instance refs. 
\cite{DE,JAP,YA}, to search for $K$-matrices solutions associated to such
quantum integrable models as well as for their elliptic extensions \cite%
{SOS,SOS1}. More recently, attempts to classify all the $K$-matrices of the
non-exceptional $U_{q}[{\cal {G}]}$ vertex models either by direct analysis
of Eq.(\ref{RE1}) \cite{LI} or by others potentially systematic approaches 
\cite{NED,CH1} have been discussed in the literature.

In spite of all these works not much is known about the structure of the $K$%
-matrices of the vertex models based on the quantum exceptional Lie
algebras. In fact, even for the simplest such case, i.e. the fundamental $%
U_{q}[G_{2}]$ $\check{R}$-matrix \cite{KU}, only the diagonal solutions have
been studied \cite{BAT1}. The purpose of this paper is to start to bridge
this gap by presenting the complete reflection matrices associated to the
minimal $U_{q}[G_{2}]$ vertex model.

This paper is organized as follows. In next section we describe the
seven-dimensional $U_{q}[G_{2}]$ $\check{R}$-matrix on the Weyl basis. This
step makes it possible to adapt the method developed for non-exceptional
models \cite{LI} to deal with the exceptional $U_{q}[G_{2}]$ case. In
section 3 we discuss what we hope to be the complete set of reflection
matrices. We find two families of non-diagonal solutions and we
confirm as well the special diagonal ones given before in ref.\cite{BAT1}. We
observe that the two types of diagonal $K$-matrices cannot be obtained as
special limits of the non-diagonal ones. Section 4 is reserved for our
conclusions. In Appendix A we present the explicit Boltzmann weights
expressions of the $U_q[G_2]$ vertex model.

\section{The $U_q[G_2]$ $R$-matrix}

In this section we shall present the $U_{q}[G_{2}]$ $\check{R}$-matrix in a
suitable basis for the analysis of the reflection equation (\ref{RE1}). In
terms of the quantum group framework \cite{JI}, this matrix can be written
as linear combination of the $U_{q}[G_{2}]$ projectors operators $%
P_{V_{\Lambda }}$ \cite{KU}: 
\begin{equation}
\check{R}(x)=\sum_{\Lambda =0,\Lambda _{1},\Lambda _{2},2\Lambda _{2}}\rho
_{\Lambda }(x)P_{V_{\Lambda }}  \label{dec}
\end{equation}%
where $V_{\Lambda }$ denotes the irreducible representations occurring in the
seven-dimensional $U_{q}[G_{2}]$ decomposition $V_{\Lambda _{2}}\otimes
V_{\Lambda _{2}}=V_{0}\oplus V_{\Lambda _{1}}\oplus V_{\Lambda _{2}}\oplus
V_{2\Lambda _{2}}$.

The weights $\rho_{\Lambda}(x)$ are functions of the quadratic Casimir
element in an irrep with the highest weight. They satisfy the following
relations, 
\begin{equation}
\rho_{\Lambda_{2}}(x)= \frac{[2]_x [12]_x}{[8]_x} \rho_{0}(x),~~~
\rho_{2\Lambda_{2}}(x)= [2]_x [12]_x \rho_{0}(x),~~~ \rho_{\Lambda_{1}}(x)=
[12]_x \rho_{0}(x)  \label{rho}
\end{equation}
where $[a]_x= \frac{x q^{a}-1}{q^a-x}$ and $\rho_{0}(x)$ is an arbitrary
normalization.

A direct analysis of the reflection equation (\ref{RE1}) is in principle
easily made by representing the $K$-matrix $K(x)$ in terms of the Weyl
basis. In this approach, the same has to be done for the $U_{q}[G_2]$ $%
\check{R}$-matrix as well. This last step involves a considerable amount of
additional work even when the projectors $P_{V_{\Lambda }}$ are known in
terms of the so-called $q$-Wigner coefficients. Ommiting here such
technicalities, we find that the $\check{R}$-matrix defined by Eqs.(\ref{dec}%
,\ref{rho}) can be rewritten in terms of the following expression,%
\begin{eqnarray}  \label{RRR}
\check{R}(x) &=&\sum_{\alpha=1, \neq 4}^{7}c_{1}(x){\rm e}_{\alpha \alpha
}\otimes {\rm e}_{\alpha \alpha }+\sum_{\overset{\alpha=1}{\alpha <4}%
}^{7}c_{3}(x)[{\rm e}_{\alpha 4}\otimes {\rm e}_{4\alpha }+{\rm e}_{4\alpha
}\otimes {\rm e}_{\alpha 4}] +\sum_{\alpha,\beta=1}^{7}c_{\alpha \beta }(x)%
{\rm e}_{\alpha ^{\prime }\beta }\otimes {\rm e}_{\alpha \beta ^{\prime }} 
\nonumber \\
&&+\sum_{\overset{\alpha,\beta=1,\neq 4}{\alpha <\beta ,\beta ^{\prime }}%
}^{7} {\cal B}_{1}(\alpha ,\beta )\left[ {\rm e}_{\alpha \alpha }\otimes 
{\rm e}_{\beta \beta }+{\rm e}_{\beta ^{\prime }\beta ^{\prime }}\otimes 
{\rm e}_{\alpha ^{\prime }\alpha ^{\prime }}\right] +{\cal A}_{1}(\alpha
,\beta )\left[ {\rm e}_{\beta \beta }\otimes {\rm e}_{\alpha \alpha }+{\rm e}%
_{\alpha ^{\prime }\alpha ^{\prime }}\otimes {\rm e}_{\beta ^{\prime }\beta
^{\prime }}\right]  \nonumber \\
&&+\sum_{\overset{\alpha=1}{\alpha <4}}^{7}{\cal B}_{2}(\alpha )\left[ {\rm e%
}_{\alpha \alpha }\otimes {\rm e}_{44}+{\rm e}_{44}\otimes {\rm e}_{\alpha
^{\prime }\alpha ^{\prime }}\right] +{\cal A}_{2}(\alpha )\left[ {\rm e}%
_{44}\otimes {\rm e}_{\alpha \alpha }+{\rm e}_{\alpha ^{\prime }\alpha
^{\prime }}\otimes {\rm e}_{44}\right]  \nonumber \\
&&+\sum_{\overset{\alpha,\beta=1,\neq 4}{\alpha <\beta ,\beta ^{\prime }}%
}^{7}{\cal C}(\alpha ,\beta )\left[ {\rm e}_{\alpha \beta }\otimes {\rm e}%
_{\beta \alpha }+{\rm e}_{\beta \alpha }\otimes {\rm e}_{\alpha \beta } + 
{\rm e}_{\alpha ^{\prime }\beta ^{\prime }}\otimes {\rm e}_{\beta ^{\prime
}\alpha ^{\prime }}+{\rm e}_{\beta ^{\prime }\alpha ^{\prime }}\otimes {\rm e%
}_{\alpha ^{\prime }\beta ^{\prime }} \right ]  \nonumber \\
&&+\sum_{k=1}^{3}\sum_{\Gamma _{k}}\left\{ {\cal B}_{[k]}\left[ {\rm e}%
_{\alpha \beta }\otimes {\rm e}_{4\delta }+{\rm e}_{\beta \alpha }\otimes 
{\rm e}_{\delta 4}+{\rm e}_{\delta ^{\prime }4}\otimes {\rm e}_{\beta
^{\prime }\alpha ^{\prime }}+{\rm e}_{4\delta ^{\prime }}\otimes {\rm e}%
_{\alpha ^{\prime }\beta ^{\prime }}\right] \right.  \nonumber \\
&&\qquad \quad +\left. {\cal A}_{[k]}\left[ {\rm e}_{\alpha ^{\prime }\beta
^{\prime }}\otimes {\rm e}_{4\delta ^{\prime }}+{\rm e}_{\beta ^{\prime
}\alpha ^{\prime }}\otimes {\rm e}_{\delta ^{\prime }4}+{\rm e}_{\delta
4}\otimes {\rm e}_{\beta \alpha }+{\rm e}_{4\delta }\otimes {\rm e}_{\alpha
\beta }\right] \right\}
\end{eqnarray}%
where $\alpha ^{\prime }=8-\alpha $ and the symbol $\Gamma _{k}$ denotes the
sum over a set of three indices $\{(\alpha ,\beta ,\delta )\}$ such that $%
\Gamma _{1}=\{(1,3,2),(5,2,7),(6,3,7)\}$, $\Gamma _{2}=(1,2,3)$ and $\Gamma
_{3}=\{(2,1,5),(3,1,6)\}$.
We also recall that
${\rm e}_{\alpha ,\beta }$ refers to the standard $7\times 7$ Weyl
matrices.

The explicit expressions of all the weights appearing in expression (\ref%
{RRR}) have been summarized in Appendix A.
An advantage of the representation (\ref{RRR}) is that it exhibits explicitly the two $U(1)$ symmetries
of the $G_2$ Lie algebra. This property will be very useful in
next section in order to classify the independent functional relations for
the reflection matrices. We also note that the last term in Eq.(\ref{RRR})
represents additional Boltzmann weights as compared with those present in
the $\check{R}(x)$-matrix of the non-exceptional vertex models \cite{JI}.

We close this section by mentioning useful relations satisfied by the matrix 
$R(x)=P\check{R}(x)$, where $P$ is the seven-dimensional permutator. Besides
the standard properties of regularity and unitarity the matrix $R(x)$
satisfies the so-called $PT$-symmetry given by 
\begin{equation}
P_{12}R_{12}(x)P_{12}=R_{12}^{t_{1}t_{2}}(x)  \label{PT}
\end{equation}%
where the symbol $t_{k}$ denotes the transposition in the space with index $%
k $.

Yet another property is the crossing symmetry, 
\begin{equation}  
\label{cross}
R_{12}(x)=\frac{-x^3}{q^{18}} V_{1} R_{12}^{t_2}(q^{12}/x )V^{-1}_{1}
\label{CRO}
\end{equation}
where $V$ is an anti-diagonal matrix whose non-null elements are $%
V_{17}=1,V_{26}=-q,V_{35}=q^4,V_{44}=-q^5,
V_{53}=q^6,V_{62}=-q^9,V_{71}=q^{10}$.

\section{The $U_q[G_2]$ $K$-matrices}

The purpose here is to search for the complete set of regular $K$-matrices
for the $U_{q}[G_{2}]$ vertex model defined in the previous section. More
specifically , we are interested in reflexion matrices having the general
form, 
\begin{equation}
K(x)=\sum_{\alpha ,\beta =1}^{7}k_{\alpha ,\beta }(x){\rm e}_{\alpha ,\beta }
\label{WEYK}
\end{equation}%
with the constraint $k_{\alpha ,\beta }(1)=\delta _{\alpha ,\beta }$. 

Direct substitution of the ansatz (\ref{WEYK}) and the $R$-matrix (\ref{RRR}%
) in Eq.(\ref{RE1}) give us several independent functional equations for
the elements $k_{\alpha ,\beta }(x)$. In order to select and solve these
equations we act on them with the operator $\frac{d}{dy}$ and afterwards we
take the $y=1$ regular limit. Here we shall denote these resulting algebraic
equations by the symbol $E[i,j]$ where the index $[i,j]$ refer to the ith
row and jth column of the original reflection equation (\ref{RE1}). Note
that such equations involve only the variable $x$ and a number of
free-parameters $\omega _{\alpha ,\beta }$ defined as 
\begin{equation}
\omega _{\alpha ,\beta }=\frac{dk_{\alpha ,\beta }(y)}{dy}|_{y=1}
\end{equation}

Our next step is to organize the algebraic equations $E[i,j]$ in suitable
blocks of relations $B[i,j]$ involving related $K$-matrices elements. We
find that such blocks are made by combining for a given pair $[i,j]$ the $%
E[i,j]$, $E[j,i]$, $E[50-i,50-j]$ and $E[50-j,50-i]$ set of relations.
Considering that we are looking for general non-diagonal $K$-matrices we
start our analysis by inspecting the blocks $B[i,j]$ possessing only
off-diagonal elements $k_{\alpha ,\beta }(x)$. The simplest ones are $%
B[1,33] $, $B[1,41]$, $B[2,42]$, $B[3,35]$, $B[9,17]$ and $B[12,20]$,
providing us constraints to the elements $k_{\alpha ,\alpha ^{\prime }}(x)$,
which reads 
\begin{equation}
k_{\alpha ,\alpha ^{\prime }}(x)=\frac{\omega _{\alpha ,a^{\prime }}}{\omega
_{1,7}}k_{1,7}(x)
\end{equation}

By the same token, the blocks $B[1,47]$, $B[1,48]$ and $B[8,38]$ lead us to
determine the following elements,%
\begin{eqnarray}
k_{1,5}(x) &=&\Gamma (x)\left( \,a_{1}\left( x\right) c_{11}\left( x\right)
\omega _{1,5}-c_{2}\left( x\right) c_{13}\left( x\right) \omega
_{3,7}\right) \frac{k_{1,7}\left( x\right) }{\omega _{1,7}} \\
k_{5,1}(x) &=&\Gamma (x){\left( \,a_{1}\left( x\right) c_{11}\left( x\right)
\omega _{5,1}-c_{2}\left( x\right) c_{13}\left( x\right) \omega
_{7,3}\right) }\frac{k_{7,1}\left( x\right) }{\omega _{7,1}} \\
k_{7,3}(x) &=&\Gamma (x){\left( b_{1}\left( x\right) c_{11}\left( x\right)
\,\omega _{7,3}-c_{2}\left( x\right) c_{31}\left( x\right) \omega
_{5,1}\right) }\frac{k_{7,1}\left( x\right) }{\omega _{7,1}} \\
k_{3,7}(x) &=&\Gamma (x){\left( b_{1}\left( x\right) c_{11}\left( x\right)
\omega _{3,7}-c_{2}\left( x\right) c_{31}\left( x\right) \omega
_{1,5}\right) }\frac{k_{1,7}\left( x\right) }{\omega _{1,7}}
\end{eqnarray}%
\begin{eqnarray}
k_{16}(x) &=&\Gamma (x)(a_{1}\left( x\right) c_{11}\left( x\right) \omega
_{1,6}-c_{2}\left( x\right) c_{12}\left( x\right) \omega _{2,7})\frac{%
k_{1,7}\left( x\right) }{\omega _{1,7}} \\
k_{61}(x) &=&\Gamma (x)(a_{1}\left( x\right) c_{11}\left( x\right) \omega
_{6,1}-c_{2}\left( x\right) c_{12}\left( x\right) \omega _{7,2})\frac{%
k_{7,1}\left( x\right) }{\omega _{7,1}} \\
k_{27}(x) &=&\Gamma (x){\left( b_{1}\left( x\right) c_{11}\left( x\right)
\,\omega _{2,7}-c_{2}\left( x\right) c_{21}\left( x\right) \omega
_{1,6}\right) }\frac{k_{1,7}\left( x\right) }{\omega _{1,7}} \\
k_{72}(x) &=&\Gamma (x){\left( b_{1}\left( x\right) c_{11}\left( x\right)
\,\omega _{7,2}-c_{2}\left( x\right) c_{21}\left( x\right) \omega
_{6,1}\right) }\frac{k_{7,1}\left( x\right) }{\omega _{7,1}}
\end{eqnarray}%
\begin{eqnarray}
k_{23}(x) &=&\Gamma (x)(a_{1}\left( x\right) c_{11}\left( x\right) \omega
_{2,3}-c_{2}\left( x\right) c_{25}\left( x\right) \omega _{5,6})\frac{%
k_{1,7}\left( x\right) }{\omega _{1,7}} \\
k_{32}(x) &=&\Gamma (x)(a_{1}\left( x\right) c_{11}\left( x\right) \omega
_{3,2}-c_{2}\left( x\right) c_{25}\left( x\right) \omega _{6,5})\frac{%
k_{7,1}\left( x\right) }{\omega _{7,1}} \\
k_{56}(x) &=&\Gamma (x){\left( b_{1}\left( x\right) c_{11}\left( x\right)
\,\omega _{5,6}-c_{2}\left( x\right) c_{52}\left( x\right) \omega
_{2,3}\right) }\frac{k_{1,7}\left( x\right) }{\omega _{1,7}} \\
k_{65}(x) &=&\Gamma (x){\left( b_{1}\left( x\right) c_{11}\left( x\right)
\,\omega _{6,5}-c_{2}\left( x\right) c_{52}\left( x\right) \omega
_{3,2}\right) }\frac{k_{7,1}\left( x\right) }{\omega _{7,1}}
\end{eqnarray}%
where we have chosen the entry $k_{1,7}(x)$ as an overall normalization. We
also have made use of the identities $%
c_{12}(x)c_{21}(x)=c_{13}(x)c_{31}(x)=c_{25}(x)c_{52}(x)$ in order to define
the common function, 
\begin{equation}
\Gamma (x)=\frac{c_{1}\left( x\right) c_{11}\left( x\right) -c_{2}\left(
x\right) c_{4}\left( x\right) }{c_{11}^{2}\left( x\right) b_{1}\left(
x\right) a_{1}\left( x\right) -c_{2}^{2}\left( x\right) c_{12}\left(
x\right) c_{21}\left( x\right) }
\end{equation}

At this point we turn our attention to the diagonal $B[i,i]$ blocks. Each of
them involves two distinct equations and after some cumbersome manipulations
we find the relation, 
\begin{equation}
\omega _{\beta ,\alpha }\ k_{\alpha ,\beta }(x)=\omega _{\alpha ,\beta }\
k_{\beta ,\alpha }(x),~~{\rm for}~~\alpha \neq \beta 
\end{equation}%
provided that the following constraints between the coefficients $\omega
_{\alpha ,\beta }$ are satisfied, 
\begin{equation}
\omega _{\alpha ,\beta }\ \omega _{\alpha ^{\prime },\beta ^{\prime
}}=\omega _{\beta ,\alpha }\ \omega _{\beta ^{\prime },\alpha ^{\prime }},~~%
{\rm for}~~\beta \neq \alpha ,\alpha ^{\prime }
\end{equation}

The last off-diagonal elements we need to determine are the entries $%
k_{4,\beta }(x)$ for $\beta >4$ and $k_{\alpha ,4}(x)$ for $\alpha <4$. It
turns out that they can be fixed by using the blocks $B[1,17]$, $B[9,33]$, $%
B[2,20]$, $B[3,12]$ and $B[8,35]$. The final results are%
\begin{eqnarray}
k_{1,4}(x) &=&\frac{\omega _{1,4}}{\omega _{2,3}}k_{2,3}(x),\ \ k_{2,4}(x)=%
\frac{\omega _{2,4}}{\omega _{1,5}}k_{1,5}(x),\ \ k_{3,4}(x)=\frac{\omega
_{3,4}}{\omega _{1,6}}k_{1,6}(x) \\
k_{4,5}(x) &=&\frac{\omega _{4,5}}{\omega _{2,7}}k_{2,7}(x),\ \ k_{4,6}(x)=%
\frac{\omega _{4,6}}{\omega _{3,7}}k_{3,7}(x),\ \ k_{4,7}(x)=\frac{\omega
_{4,7}}{\omega _{5,6}}k_{5,6}(x)
\end{eqnarray}

We now reached a point in which all the above considerations can be
collected together. Substituting the determined off-diagonal elements back
to the reflection equation (\ref{RE1}) we find out that the parameter $%
\omega _{2,7}$ has to satisfy the following polynomial equation, 
\begin{equation}
\omega _{2,7}(\omega _{2,7}-\omega _{1,6}q^{-5})(\omega _{2,7}+\omega
_{1,6}q^{-5})(\omega _{2,7}+\omega _{1,6}q^{-8})=0
\end{equation}

The values $\omega _{2,7}=0,-\omega _{1,6}q^{-8}$ lead us to solutions
whose non-null elements are only the diagonal $k_{\alpha ,\alpha }(x)$ and
the anti-diagonal $k_{\alpha ,\alpha ^{\prime }}$ entries. This branch gives
origin to two distinct classes of solution depending on whether $\omega
_{1,7}=0$ or $\omega _{1,7}\neq 0$. By setting $\omega _{1,7}=0$ we get the
two diagonal solutions found previously in ref.\cite{BAT1}, which in current
notation reads, 
\begin{eqnarray}
K^{(1)}(x) &=& I_{7}~~~~ \\
&&  \nonumber \\
K^{(2)}(x) &=&{\rm Diag}(1,1,x\frac{qx+\epsilon }{q+\epsilon x},x\frac{%
qx+\epsilon }{q+\epsilon x},x\frac{qx+\epsilon }{q+\epsilon x},x^{2},x^{2})
\end{eqnarray}%
where $\epsilon= \pm 1 $ is a discrete parameter. 

On the other hand, for $\omega _{1,7}\neq 0$ we obtain the following novel
non-diagonal solution 
\begin{equation}
K^{(3)}(x)=\left( 
\begin{array}{ccccccc}
k_{1,1}^{(3)} &  &  &  &  &  & k_{1,7}^{(3)}(x) \\ 
& k_{2,2}^{(3)} &  &  &  & k_{2,6}^{(3)} &  \\ 
&  & k_{3,3}^{(3)} &  & k_{3,5}^{(3)} &  &  \\ 
&  &  & k_{4,4}^{(3)} &  &  &  \\ 
&  & k_{5,3}^{(3)} &  & k_{5,5}^{(3)} &  &  \\ 
& k_{6,2}^{(3)} &  &  &  & k_{6,6}^{(3)} &  \\ 
k_{7,1}^{(3)} &  &  &  &  &  & k_{7,7}^{(3)}%
\end{array}%
\right)
\end{equation}%
where the diagonal entries are given by, 
\begin{eqnarray}
k_{1,1}^{(3)} &=&k_{2,2}^{(3)}=k_{3,3}^{(3)}=\frac{1}{\omega _{1,7}}\frac{2}{%
x^{2}-1}k_{1,7}^{(3)}(x) \\
k_{4,4}^{(3)} &=&\frac{1}{\omega _{1,7}}\frac{q^{2}-x^{2}}{q^{2}-1}\frac{2}{%
x^{2}-1}k_{1,7}^{(3)}(x) \\
k_{5,5}^{(3)} &=&k_{6,6}^{(3)}=k_{7,7}^{(3)}=\frac{1}{\omega _{1,7}}\frac{%
2x^{2}}{x^{2}-1}k_{1,7}^{(3)}(x)
\end{eqnarray}%
while the off-diagonal elements are, 
\begin{eqnarray}
k_{2,6}^{(3)} &=&\frac{\omega _{2,6}}{\omega _{1,7}}k_{1,7}^{(3)}(x) \\
k_{3,5}^{(3)} &=&-\frac{1}{\omega _{2,6}}\left( \frac{2q}{q^{2}-1}\right)
k_{1,7}^{(3)}(x) \\
k_{5,3}^{(3)} &=&-\frac{\omega _{2,6}}{\omega _{1,7}^{2}}\left( \frac{2q}{%
q^{2}-1}\right) k_{1,7}^{(3)}(x) \\
k_{6,2}^{(3)} &=&\frac{1}{\omega _{1,7}\omega _{2,6}}\left( \frac{2q}{q^{2}-1%
}\right) ^{2}k_{1,7}^{(3)}(x) \\
k_{7,1}^{(3)} &=&\frac{1}{\omega _{1,7}^{2}}\left( \frac{2q}{q^{2}-1}\right)
^{2}k_{1,7}^{(3)}(x)
\end{eqnarray}

The remaining branch $\omega _{2,7}=\epsilon \omega _{1,6}q^{-5}$ is the
most complicated one since it leads us to $K$-matrices with all non-null
elements. Denoting the entries of this complete matrix by $k^{(4)}{}_{\alpha
,\beta }(x)$ we find, after elaborated algebraic manipulations, that their
expressions can be defined as follows. The diagonal entries are, 
\begin{equation}
k_{1,1}^{(4)}(x)={\cal F}^{(\epsilon )}\left( 2,2\right) \left( \frac{{q}%
^{10}\left( {q}^{8}-1\right) \left( 1-\epsilon x\right) }{\left(
x^{2}-1\right) \left( {q}^{5}-\epsilon x\right) }-\frac{q^{9}\left(
1+q\right) \left( {q}^{8}+\epsilon \,x\right) \left( 1+{q}^{4}\right) }{%
\left( {q}^{3}+1\right) \left( x^{2}-1\right) \left( {q}^{5}-\epsilon
x\right) }\right) k_{1,7}^{(4)}\left( x\right) 
\end{equation}%
\begin{equation}
k_{2,2}^{(4)}(x)=k_{1,1}^{(4)}\left( x\right) -{\cal F}^{(\epsilon )}\left(
2,2\right) {\frac{q^{10}\left( {q}^{3}+1\right) }{{q}^{5}-\epsilon x}}%
k_{1,7}^{(4)}\left( x\right) 
\end{equation}%
\begin{equation}
k_{3,3}^{(4)}(x)=k_{1,1}^{(4)}\left( x\right) +{\cal F}^{(\epsilon )}\left(
2,2\right) {\frac{q^{10}\left( {q}^{6}-1\right) }{{q}^{5}-\epsilon x}}%
k_{1,7}^{(4)}\left( x\right) 
\end{equation}%
\begin{equation}
k_{4,4}^{(4)}(x)={\cal F}^{(\epsilon )}\left( 2,2\right) \left( \frac{{q}%
^{12}\left( {q}^{2}-x^{2}\right) }{\left( {q}^{3}+1\right) \left(
x^{2}-1\right) }-{\frac{q^{9}\left( {q}^{9}+1\right) \left( \epsilon
\,qx+1\right) }{\left( x^{2}-1\right) \left( {q}^{5}-\epsilon \,x\right) }}%
\epsilon \,x\right) k_{1,7}^{(4)}\left( x\right) 
\end{equation}%
\begin{equation}
k_{5,5}^{(4)}(x)=x^{2}k_{1,1}^{(4)}\left( x\right) +{\cal F}^{(\epsilon
)}\left( 2,2\right) {\frac{{q}^{12}\left( {q}^{6}-1\right) }{{q}%
^{5}-\epsilon \,x}}\epsilon \,xk_{1,7}^{(4)}\left( x\right) 
\end{equation}%
\begin{equation}
k_{6,6}^{(4)}(u)=x^{2}k_{1,1}^{(4)}\left( x\right) +{\cal F}^{(\epsilon
)}\left( 2,2\right) {\frac{\,q^{15}\left( {q}^{3}+1\right) }{{q}%
^{5}-\epsilon \,x}}\epsilon \,xk_{1,7}^{(k)}\left( x\right) 
\end{equation}%
\begin{equation}
k_{7,7}^{(4)}(x)=x^{2}k_{1,1}^{(4)}\left( x\right) 
\end{equation}

The structure of the off-diagonal entries shall be described in terms of the
elements of each row separately. For the first row we find,
\begin{eqnarray}
k_{12}^{\left( 4\right) }(x) &=&-{\cal F}^{(\epsilon )}\left( 1,2\right) {%
\frac{{q}^{9}}{{q}^{5}-\epsilon \,x}\ }k_{1,7}^{\left( 4\right) }\left(
x\right) , \\
k_{13}^{\left( 4\right) }(x) &=&-{\cal F}^{(\epsilon )}\left( 2,1\right) {%
\frac{{q}^{9}}{{q}^{5}-\epsilon \,x}\ }k_{1,7}^{\left( 4\right) }\left(
x\right) , \\
k_{14}^{\left( 4\right) }(x) &=&\sqrt{q^{2}\left( 1+{q}^{2}\right) }{\cal F}%
^{(\epsilon )}\left( 1,1\right) \frac{{q}^{6}}{{q}^{5}-\epsilon \,x}\
k_{1,7}^{\left( 4\right) }\left( x\right) , \\
k_{15}^{\left( 4\right) }(x) &=&-{\cal F}^{(\epsilon )}\left( 0,1\right) {%
\frac{{q}^{6}}{{q}^{5}-\epsilon \,x}\ }k_{1,7}^{\left( 4\right) }\left(
x\right) , \\
k_{16}^{\left( 4\right) }(x) &=&-{\cal F}^{(\epsilon )}\left( 1,0\right) {%
\frac{{q}^{6}}{{q}^{5}-\epsilon \,x}\ }k_{1,7}^{\left( 4\right) }\left(
x\right) .
\end{eqnarray}%
The second row is,

\begin{eqnarray}
k_{21}^{\left( 4\right) }(x) &=&{\cal F}^{(\epsilon )}\left( 3,2\right) {%
\frac{{q}^{14}}{{q}^{5}-\epsilon \,x}\ }k_{1,7}^{\left( 4\right) }\left(
x\right) , \\
k_{23}^{\left( 4\right) }(x) &=&-{\cal F}^{(\epsilon )}\left( 3,1\right) {%
\frac{{q}^{13}}{{q}^{5}-\epsilon \,x}\ }k_{1,7}^{\left( 4\right) }\left(
x\right) , \\
k_{24}^{\left( 4\right) }(x) &=&\sqrt{q^{2}\left( 1+{q}^{2}\right) }{\cal F}%
^{(\epsilon )}\left( 2,1\right) \frac{{q}^{10}}{{q}^{5}-\epsilon \,x}\
k_{1,7}^{\left( 4\right) }\left( x\right) , \\
k_{25}^{\left( 4\right) }(x) &=&-{\cal F}^{(\epsilon )}\left( 1,1\right) {%
\frac{{q}^{10}}{{q}^{5}-\epsilon \,x}\ }k_{1,7}^{\left( 4\right) }\left(
x\right) , \\
k_{26}^{\left( 4\right) }(x) &=&-{\cal F}^{(\epsilon )}\left( 2,0\right) {%
\frac{{q}^{5}}{{q}^{3}+1}\ }k_{1,7}^{\left( 4\right) }\left( x\right) , \\
k_{27}^{\left( 4\right) }(x) &=&-{\cal F}^{(\epsilon )}\left( 1,0\right) {%
\frac{\,q}{{q}^{5}-\epsilon \,x}\ \epsilon }xk_{1,7}^{\left( 4\right)
}\left( x\right) .
\end{eqnarray}%
The third row is,

\begin{eqnarray}
k_{31}^{\left( 4\right) }(x) &=&-{\cal F}^{(\epsilon )}\left( 2,3\right) {%
\frac{{q}^{17}}{{q}^{5}-\epsilon \,x}}k_{1,7}^{\left( 4\right) }\left(
x\right) , \\
k_{32}^{\left( 4\right) }(x) &=&{\cal F}^{(\epsilon )}\left( 1,3\right) {%
\frac{{q}^{16}}{{q}^{5}-\epsilon \,x}}k_{1,7}^{\left( 4\right) }\left(
x\right) , \\
k_{34}^{\left( 4\right) }(x) &=&-\sqrt{q^{2}\left( 1+{q}^{2}\right) }{\cal F}%
^{(\epsilon )}\left( 1,2\right) \frac{{q}^{13}}{{q}^{5}-\epsilon \,x}%
k_{1,7}^{\left( 4\right) }\left( x\right) , \\
k_{35}^{\left( 4\right) }(x) &=&{\cal F}^{(\epsilon )}\left( 0,2\right) {%
\frac{{q}^{8}}{{q}^{3}+1}}k_{1,7}^{\left( 4\right) }\left( x\right) , \\
k_{36}^{\left( 4\right) }(x) &=&-{\cal F}^{(\epsilon )}\left( 1,1\right) {%
\frac{\,{q}^{5}}{{q}^{5}-\epsilon \,x}\ \epsilon }xk_{1,7}^{\left( 4\right)
}\left( x\right) , \\
k_{37}^{\left( 4\right) }(x) &=&{\cal F}^{(\epsilon )}\left( 0,1\right) {%
\frac{{q}^{4}}{{q}^{5}-\epsilon \,x}\ \epsilon }xk_{1,7}^{\left( 4\right)
}\left( x\right) .
\end{eqnarray}%
The fourth row is,

\begin{eqnarray}
k_{41}^{\left( 4\right) }(x) &=&-\sqrt{q^{-2}\left( 1+{q}^{2}\right) }{\cal F%
}^{(\epsilon )}\left( 3,3\right) \frac{{q}^{21}}{{q}^{5}-\epsilon \,x}\
k_{1,7}^{\left( 4\right) }\left( x\right) , \\
k_{42}^{\left( 4\right) }(x) &=&\sqrt{q^{2}\left( 1+{q}^{2}\right) }{\cal F}%
^{(\epsilon )}\left( 2,3\right) \frac{{q}^{18}}{{q}^{5}-\epsilon \,x}\
k_{1,7}^{\left( 4\right) }\left( x\right) , \\
k_{43}^{\left( 4\right) }(x) &=&\sqrt{q^{2}\left( 1+{q}^{2}\right) }{\cal F}%
^{(\epsilon )}\left( 3,2\right) \frac{{q}^{18}}{{q}^{5}-\epsilon \,x}\
k_{1,7}^{\left( 4\right) }\left( x\right) , \\
k_{45}^{\left( 4\right) }(x) &=&-\sqrt{q^{2}\left( 1+{q}^{2}\right) }{\cal F}%
^{(\epsilon )}\left( 1,2\right) \frac{{q}^{7}}{{q}^{5}-\epsilon \,x}\
\epsilon \,xk_{1,7}^{\left( 4\right) }\left( x\right) , \\
k_{46}^{\left( 4\right) }(x) &=&-\sqrt{q^{2}\left( 1+{q}^{2}\right) }{\cal F}%
^{(\epsilon )}\left( 2,1\right) \frac{{q}^{7}}{{q}^{5}-\epsilon \,x}\
\epsilon \,xk_{1,7}^{\left( 4\right) }\left( x\right) , \\
k_{47}^{\left( 4\right) }(x) &=&\sqrt{q^{2}\left( 1+{q}^{2}\right) }{\cal F}%
^{(\epsilon )}\left( 1,1\right) \frac{{q}^{6}}{{q}^{5}-\epsilon \,x}\
\epsilon \,xk_{1,7}^{\left( 4\right) }\left( x\right) .
\end{eqnarray}%
The fifth row is,

\begin{eqnarray}
k_{51}^{\left( 4\right) }(x) &=&-{\cal F}^{(\epsilon )}\left( 4,3\right) {%
\frac{{q}^{24}}{{q}^{5}-\epsilon \,x}\ }k_{1,7}^{\left( 4\right) }\left(
x\right) , \\
k_{52}^{\left( 4\right) }(x) &=&{\cal F}^{(\epsilon )}\left( 3,3\right) {%
\frac{{q}^{23}}{{q}^{5}-\epsilon \,x}\ }k_{1,7}^{\left( 4\right) }\left(
x\right) , \\
k_{53}^{\left( 4\right) }(x) &=&{\cal F}^{(\epsilon )}\left( 4,2\right) {%
\frac{{q}^{18}}{{q}^{3}+1}\ }k_{1,7}^{\left( 4\right) }\left( x\right) , \\
k_{54}^{\left( 4\right) }(x) &=&\sqrt{\left( 1+{q}^{2}\right) q^{2}}{\cal F}%
^{(\epsilon )}\left( 3,2\right) \frac{{q}^{12}}{{q}^{5}-\epsilon \,x}\
\epsilon \,xk_{1,7}^{\left( 4\right) }\left( x\right) , \\
k_{56}^{\left( 4\right) }(x) &=&-{\cal F}^{(\epsilon )}\left( 3,1\right) {%
\frac{{q}^{12}}{{q}^{5}-\epsilon \,x}\ }\epsilon \,xk_{1,7}^{\left( 4\right)
}\left( x\right) , \\
k_{57}^{\left( 4\right) }(x) &=&{\cal F}^{(\epsilon )}\left( 2,1\right) {%
\frac{{q}^{11}}{{q}^{5}-\epsilon \,x}\ }\epsilon \,xk_{1,7}^{\left( 4\right)
}\left( x\right) .
\end{eqnarray}%
The sixth row is,

\begin{eqnarray}
k_{61}^{\left( 4\right) }(x) &=&{\cal F}^{(\epsilon )}\left( 3,4\right) {%
\frac{{q}^{27}}{{q}^{5}-\epsilon \,x}\ }k_{1,7}^{\left( 4\right) }\left(
x\right) , \\
k_{62}^{\left( 4\right) }(x) &=&-{\cal F}^{(\epsilon )}\left( 2,4\right) {%
\frac{{q}^{21}}{{q}^{3}+1}\ }k_{1,7}^{\left( 4\right) }\left( x\right) , \\
k_{63}^{\left( 4\right) }(x) &=&{\cal F}^{(\epsilon )}\left( 3,3\right) {%
\frac{{q}^{18}}{{q}^{5}-\epsilon \,x}\ }\epsilon \,xk_{1,7}^{\left( 4\right)
}\left( x\right) , \\
k_{64}^{\left( 4\right) }(x) &=&-\sqrt{\left( 1+{q}^{2}\right) q^{2}}{\cal F}%
^{(\epsilon )}\left( 2,3\right) \frac{{q}^{15}}{{q}^{5}-\epsilon \,x}\
\epsilon \,xk_{1,7}^{\left( 4\right) }\left( x\right) , \\
k_{65}^{\left( 4\right) }(x) &=&{\cal F}^{(\epsilon )}\left( 1,3\right) {%
\frac{{q}^{15}}{{q}^{5}-\epsilon \,x}\ }\epsilon \,xk_{1,7}^{\left( 4\right)
}\left( x\right) , \\
k_{67}^{\left( 4\right) }(x) &=&-{\cal F}^{(\epsilon )}\left( 1,2\right) {%
\frac{{q}^{14}}{{q}^{5}-\epsilon \,x}\ }\epsilon \,xk_{1,7}^{\left( 4\right)
}\left( x\right) ,
\end{eqnarray}%
The seventh row is,

\begin{eqnarray}
k_{71}^{\left( 4\right) }(x) &=&{\cal F}^{(\epsilon )}\left( 4,4\right) {%
\frac{{q}^{26}}{{q}^{3}+1}\ }k_{1,7}^{\left( 4\right) }\left( x\right) , \\
k_{72}^{\left( 4\right) }(x) &=&{\cal F}^{(\epsilon )}\left( 3,4\right) {%
\frac{{q}^{22}}{{q}^{5}-\epsilon \,x}\ }\epsilon \,xk_{1,7}^{\left( 4\right)
}\left( x\right) , \\
k_{73}^{\left( 4\right) }(x) &=&{\cal F}^{(\epsilon )}\left( 4,3\right) {%
\frac{{q}^{22}}{{q}^{5}-\epsilon \,x}\ }\epsilon \,xk_{1,7}^{\left( 4\right)
}\left( x\right) , \\
k_{74}^{\left( 4\right) }(x) &=&-\sqrt{q^{-2}{(1+{q}^{2})}}{\cal F}%
^{(\epsilon )}\left( 3,3\right) \frac{{q}^{21}}{{q}^{5}-\epsilon \,x}\
\epsilon \,xk_{1,7}^{\left( 4\right) }\left( x\right) , \\
k_{75}^{\left( 4\right) }(x) &=&{\cal F}^{(\epsilon )}\left( 2,3\right) {%
\frac{{q}^{19}}{{q}^{5}-\epsilon \,x}\ }\epsilon \,xk_{1,7}^{\left( 4\right)
}\left( x\right) , \\
k_{76}^{\left( 4\right) }(x) &=&{\cal F}^{(\epsilon )}\left( 3,2\right) {%
\frac{{q}^{19}}{{q}^{5}-\epsilon \,x}\ }\epsilon \,xk_{1,7}^{\left( 4\right)
}\left( x\right) ,
\end{eqnarray}%
where the auxiliary function ${\cal F}^{(\epsilon )}(n,m)$ is defined by, 
\begin{equation}
{\cal F}^{(\epsilon )}(n,m)={\frac{{2}^{n+m}\left( {q}^{3}+1\right) ^{n+m+1}%
}{\left( \epsilon +q\right) ^{n+m}\left( \epsilon +{q}^{4}\right)
^{n+m}\left( \epsilon +{q}^{8}\right) ^{n+m}\omega _{1,2}^{n}\omega _{1,3}^{m}}%
.}
\end{equation}

Note that this latter solution contains an additional discrete parameter besides
the two continuous  $\omega_{1,2}$ and $\omega_{1,3}$ variables. 
We also observe that a striking feature of our
results is that the non-diagonal solutions do not reduce to any of the
admissible diagonal matrices $K^{(1)}(x)$ and $K^{(2)}(x)$. This means that
the four possibilities discussed above are indeed the distinct types of $K$%
-matrices of the $U_{q}[G_{2}]$ model. The latter feature should be
contrasted with the findings for the non-exceptional vertex models \cite{LI}
in which reductions of general non-diagonal matrices guide us always to some
particular diagonal solution. It remains to be seen whether such peculiarity
is special to the $U_{q}[G_{2}]$ symmetry or is valid for all exceptional
vertex model as well.  

Finally, 
equipped with the four reflection matrices $K^{(l)}(x)$,
an integrable model with open boundary condition
can be obtained through the double-row transfer matrix
formulated by Sklyanin \cite{SK}, namely
\begin{equation}
\label{mode}
t^{(l,m)}(x) ={\rm Tr}_a\left[\stackrel{a}{K}_+^{(m)}(x)
T(x )\stackrel{a}{K}^{(l)}(x)T^{-1}(1/x)\right]~~~{\rm for}~~~l,m=1,\cdots,4
\end{equation}
where $T(x)=R_{aL}(x) \cdots R_{a1}(x)$ 
is the standard monodromy matrix of the 
corresponding closed chain with $L$ sites.  
The matrix $K^{(m)}_{+}(x)$ is
automatically determined \cite{NEP} from 
$K^{(m)}(x)$ with the help of the 
crossing property (\ref{cross}),
\begin{equation}
K_{+}^{(m)}(x)=
\left[ K^{(m)}(q^{12}/x) \right]^{t} V^{t} V 
\end{equation}

\section{Conclusions}

In this paper we have been able to classify the possible families of
reflexion matrices associated to the fundamental $U_q[G_2]$ vertex model. We
find that there exists four different classes of $K$-matrices. Two of them
are the diagonal solutions studied before \cite{BAT1} and the remaining ones
are new non-diagonal $K$-matrices with continuous free parameters. The first
type of non-diagonal solution is made by combining only the diagonal and the
anti-diagonal elements while the second one consists of matrix whose all
entries are non-null and has an extra discrete parameter. Interesting
enough, these non-diagonal solutions do not possess reduction to particular
diagonal matrices for any values of the free parameters at disposal. We
stress that this peculiarity does not occur in the $K$-matrices of the
vertex models based on the non- exceptional Lie algebras \cite{LI}.

We hope that our results will prompt further lines of investigations. One
possibility is to explore the Weyl representation of the $U_q[G_2]$ $\check{R%
}$-matrix to establish, at least for diagonal boundaries, the diagonalization 
of the transfer-matrix (\ref{mode})
from a first principle framework such as the quantum inverse
scattering method \cite{KOR}. This will give us an opportunity to check
certain assumptions, usually denominated ``doubling hypotheses'' \cite{ART},
used to determine the corresponding Bethe ansatz
solution \cite{BAT1}. 
.

\section*{Acknowledgements}

The work of A. Lima-Santos and M.J. Martins has been partially supported by
the Brazilian research Agencies CNPq and Fapesp.

\addcontentsline{toc}{section}{Appendix A}

\section*{\bf Appendix A: The Boltzmann weights}

\setcounter{equation}{0} \renewcommand{\theequation}{A.\arabic{equation}}

In this Appendix we are going to describe all the Boltzmann weights
appearing in Eq.(\ref{RRR}). In order to do that we find convenient to
introduce an auxiliary function $\varphi (\alpha ,\beta )$ defined by, 
\begin{equation}
\varphi (\alpha ,\beta )=\left\{ 
\begin{array}{c}
\lbrack \frac{\beta }{2}]-\alpha +4\delta _{4,\beta }\qquad \quad {\rm for\ }%
\alpha <\beta ,\ \alpha <\beta ^{\prime } \\ 
\lbrack \frac{\alpha ^{\prime }}{2}]-\beta ^{\prime }+4\delta _{\alpha
^{\prime },4}\qquad {\rm for\ }\alpha <\beta ,\ \alpha >\beta ^{\prime } \\ 
\beta -[\frac{\alpha }{2}]-4\delta _{\alpha ,4}\qquad \quad {\rm for\ }%
\alpha >\beta ,\ \alpha ^{\prime }>\beta \\ 
\alpha ^{\prime }-[\frac{\beta ^{\prime }}{2}]-4\delta _{4,\beta ^{\prime
}}\qquad {\rm for\ }\alpha >\beta ,\ \alpha ^{\prime }<\beta%
\end{array}%
\right.
\end{equation}
where $[\frac{\beta }{2}]$ denotes the integer part of $\frac{\beta }{2}$.

We begin by listing the weights ${\cal A}_{1}(\alpha ,\beta )$, ${\cal A}%
_{2}(\alpha)$ and ${\cal A}_{[k]}$ as well as ${\cal B}_{1}(\alpha ,\beta )$%
, ${\cal B}_{2}(\alpha)$ and ${\cal B}_{[k]}$. They are given by,

\begin{eqnarray}
{\cal A}_{1}(\alpha ,\beta ) &=&\left\{ 
\begin{array}{c}
a_{7}(x)\quad {\rm for}\ \varphi (\alpha ,\beta )=-1 \\ 
a_{1}(x)\quad {\rm for}\ \varphi (\alpha ,\beta )=0\ \  \\ 
a_{6}(x)\quad {\rm for}\ \varphi (\alpha ,\beta )=1,2%
\end{array}%
\right. ,{\cal A}_{2}(\alpha )=\left\{ 
\begin{array}{c}
a_{4}(x)\quad {\rm for}\ \varphi (\alpha ,4 )=5 \\ 
a_{8}(x)\quad {\rm for}\ \varphi (\alpha ,4 )\neq 5%
\end{array}%
\right. , \\
{\cal A}_{[1]} &=&a_{2}(x),\qquad {\cal A}_{[2]}=a_{3}(x),\qquad {\cal A}%
_{[3]}=a_{5}(x)
\end{eqnarray}%
\begin{eqnarray}
{\cal B}_{1}(\alpha ,\beta ) &=&\left\{ 
\begin{array}{c}
b_{7}(x)\quad {\rm for}\ \varphi (\alpha ,\beta )=-1 \\ 
b_{1}(x)\quad {\rm for}\ \varphi (\alpha ,\beta )=0\ \  \\ 
b_{6}(x)\quad {\rm for}\ \varphi (\alpha ,\beta )=1,2%
\end{array}%
\right. ,{\cal B}_{2}(\alpha )=\left\{ 
\begin{array}{c}
b_{4}(x)\quad {\rm for}\ \varphi (\alpha ,4)=5 \\ 
b_{8}(x)\quad {\rm for}\ \varphi (\alpha ,4)\neq 5%
\end{array}%
\right. , \\
{\cal B}_{[1]} &=&b_{2}(x),\qquad {\cal B}_{[2]}=b_{3}(x),\qquad {\cal B}%
_{[3]}=b_{5}(x)
\end{eqnarray}%
while the weight ${\cal C}(\alpha ,\beta )$ is, 
\begin{equation}
{\cal C}(\alpha ,\beta )=\left\{ 
\begin{array}{c}
c_{2}(x)\qquad {\rm for}\ \varphi (\alpha ,\beta )=0 \\ 
c_{4}(x)\qquad {\rm for}\ \varphi (\alpha ,\beta )\neq 0%
\end{array}%
\right. ,
\end{equation}

The functions $a_{\alpha }(x)$ and $b_{\alpha }(x)$ entering in the above
expressions are, 
\begin{eqnarray}
a_{1}(u) &=&\left( {q}^{2}-1\right) \left( {q}^{8}-x\right) \left( {q}%
^{12}-x\right) ,\quad b_{1}(x)=x{a}_{1}(x),\quad  \\
a_{2}(u) &=&-q^{7}\sqrt{1+{q}^{-2}}\left( {q}^{2}-1\right) \left( {q}%
^{12}-x\right) \left( x-1\right) ,\quad b_{2}(x)=\frac{x}{q^{5}}a_{2}(x), \\
a_{3}(u) &=&q^{4}\sqrt{1+{q}^{-2}}\left( {q}^{2}-1\right) \left( {q}%
^{12}-x\right) \left( x-1\right) ,\quad b_{3}(x)=xqa_{3}(x),\quad  \\
a_{5}(u) &=&q^{5}\sqrt{1+{q}^{-2}}\left( {q}^{2}-1\right) \left( {q}%
^{12}-x\right) \left( x-1\right) ,\quad b_{5}(x)=\frac{x}{q}a_{5}(x), \\
a_{4}(x) &=&\left( {q}^{4}-1\right) \left( {q}^{12}-x\right) f_{1}(x),\quad
b_{4}(x)=\frac{xf_{2}(x)}{f_{1}(x)}{a}_{4}(x), \\
a_{6}(x) &=&\left( q^{2}-1\right) \left( {q}^{12}-x\right) g_{1}(x),\quad
b_{6}(x)=\frac{xg_{2}(x)}{g_{1}(x)}{a}_{6}(x), \\
a_{7}(x) &=&\left( q^{2}-1\right) \left( {q}^{12}-x\right) g_{2}(x),\quad
b_{7}(x)=\frac{xg_{1}(x)}{g_{2}(x)}a_{7}(x), \\
a_{8}(x) &=&\left( {q}^{4}-1\right) \left( {q}^{12}-x\right) f_{2}(x),\quad
b_{8}(x)=\frac{xf_{1}(x)}{f_{2}(x)}a_{8}(x),
\end{eqnarray}%
where 
\begin{eqnarray}
f_{1}(x) &=&{q}^{2}(1-{q}^{2}+{q}^{4})-x, \\
f_{2}(x) &=&{q}^{6}-\left( 1-{q}^{2}+{q}^{4}\right) x, \\
g_{1}(x) &=&{q}^{2}(1+{q}^{2}+{q}^{6})-\left( 1+{q}^{2}+{q}^{4}\right) x, \\
g_{2}(x) &=&{q}^{4}(1+{q}^{2}+{q}^{4})-\left( 1+{q}^{4}+{q}^{6}\right) x.
\end{eqnarray}

Next, the weights $c_{\alpha }(x)$ are, 
\begin{eqnarray}
c_{1}(x) &=&(q^{2}-x)(q^{8}-x)(q^{12}-x), \\
c_{2}(x) &=&-q(q^{8}-x)(q^{12}-x)(x-1), \\
c_{3}(x) &=&-q^{2}(q^{6}-x)(q^{12}-x)(x-1), \\
c_{4}(x) &=&-q^{3}(q^{4}-x)(q^{12}-x)(x-1)
\end{eqnarray}%
while the $c_{\alpha \beta }(x)$ weights are given by 
\begin{equation}
c_{\alpha \beta }(x)=\left\{ 
\begin{array}{c}
-{q}^{4}\left( {q}^{4}-x\right) \left( {q}^{10}-x\right) \left( x-1\right)
,\quad (\alpha =\beta ,\alpha \neq 4) \\ 
\left( {q}^{6}-x\right) [{x}^{2}+\left( 1+{q}^{4}\right) [\left( {q}%
^{4}-1\right) ^{2}\left( {q}^{2}-1\right) ^{2}-{q}^{6}]x+{q}^{14}],\quad
(\alpha =\beta =4)\quad  \\ 
(-1)^{\alpha +\beta }(q^{2}-1)q^{16+\overset{\_}{\alpha }-\overset{\_}{\beta 
}}(x-1){\cal F}_{\varphi (\alpha ,\beta )},\qquad (\alpha <\beta ,\quad
\beta \neq \alpha ^{\prime }) \\ 
(-1)^{\alpha +\beta }(q^{2}-1)q^{\overset{\_}{\alpha }-\overset{\_}{\beta }%
}(x-1)x{\cal F}_{\varphi (\alpha ,\beta )},\qquad (\alpha >\beta ,\quad
\beta \neq \alpha ^{\prime })%
\end{array}%
\right. 
\end{equation}%
where the auxiliary index $\overset{\_}{\alpha }$ is defined as, 
\begin{equation}
\overset{\_}{1}=0,\ \overset{\_}{2}=3,\ \overset{\_}{3}=6,\ \overset{\_}{4}%
=8,\ \overset{\_}{5}=10,\ \overset{\_}{6}=13,\ \overset{\_}{7}=16
\end{equation}%
and the functions ${\cal F}_{\varphi (\alpha ,\beta )}{\cal \ }$are 
\begin{eqnarray}
{\cal F}_{0} &=&q^{4}-x \\
{\cal F}_{5} &=&{\cal F}_{-3}={\cal F}_{-4}=\frac{1}{q}{\left( 1+{q}%
^{2}\right) [q^{6}(1-q}^{2}+{{q}^{4})-x]} \\
{\cal F}_{-5} &=&{\cal F}_{3}={\cal F}_{4}=\frac{1}{q^{7}}{\left( 1+{q}%
^{2}\right) [{q}^{10}-(1-{q}^{2}+{\ q}^{4})x]} \\
{\cal F}_{1} &=&{\cal F}_{2}=\frac{1}{q^{2}}[{{q}^{8}\left( 1+{q}^{4}+{q}%
^{6}\right) -\left( 1+{q}^{2}+{q}^{4}\right) x]} \\
{\cal F}_{-1} &=&{\cal F}_{-2}=\frac{1}{q^{8}}[{{q}^{10}\left( 1+{q}^{2}+{q}%
^{4}\right) -\left( 1+{q}^{2}+{q}^{6}\right) x]}
\end{eqnarray}

Finally, the remaining weights $c_{\alpha, \alpha ^{\prime }}(x)$ are given
by 
\begin{eqnarray}
c_{17}(x) &=&(q^{2}-1)(q^{8}-1)[q^{8}(1+q^{4})-(q^{8}+1)x] \\
c_{26}(x)
&=&(q^{2}-1)[q^{10}(q^{10}-1)-q^{6}(q^{2}-1)(q^{4}+1)x-(q^{6}-1)x^{2}] \\
c_{35}(x) &=&(q^{2}-1)[q^{16}(q^{4}-1)+q^{8}(q^{8}-1)x-(q^{12}-1)x^{2}] \\
c_{53}(x) &=&(q^{2}-1)[q^{8}(q^{12}-1)-q^{4}(q^{8}-1)x-(q^{4}-1)x^{2}]x \\
c_{62}(x)
&=&(q^{2}-1)[q^{14}(q^{6}-1)+q^{8}(q^{2}-1)(q^{4}+1)x-(q^{10}-1)x^{2}]x \\
c_{71}(x) &=&(q^{2}-1)(q^{8}-1)[q^{4}(q^{8}+1)-(q^{4}+1)x]x^{2}
\end{eqnarray}

\end{document}